\documentclass[9pt,twocolumn,twoside]{osajnl}
\graphicspath{{.}{./figs/}}
\journal{ol} 

\setboolean{shortarticle}{true}

\title{ Inverse design of an all-dielectric digitized terahertz optical power splitter by particle swarm optimization method}

\author[*]{Kiyanoush Goudarzi and Haewook Han}

\affil[ ]{Department of Electrical Engineering, Pohang University of Science and Technology, Pohang 37673, Republic of Korea}

\affil[*]{Corresponding author: hhan@postech.ac.kr}

\begin{abstract}
This letter proposes a 50:50 TE terahertz (THz) power splitter based on digitized metamaterials. The device is optimized using particle swarm optimization algorithm. The structure of the digitized metamaterials contains circular air holes in silicon slab which are deposited on silicon dioxide substrate. The optimized structure shows insertion loss (IL) of less than 3.9 dB over the frequency range of 0.9--1.1 THz. The proposed device has the footprint of 400 $\times$ 350 \textmu m\textsuperscript{2}. The proposed device has excellent merits such as compactness, efficient, all-dielectric, and compatible with CMOS fabrication technologies. This device can be used as a vital component in most of the THz circuits to route THz waves.
\end{abstract}

\setboolean{displaycopyright}{true}

\begin{document}

\maketitle\\

\noindent THz technology has aroused extensive attention of many researchers due to the wonderful properties of THz waves. These waves are applicable in medicine \cite{siegel2004terahertz,brun2010terahertz,smye2001interaction}, security \cite{ergun2015terahertz,cheng2020concealed}, communication \cite{harter2020generalized,sarieddeen2020next,petrov2020capacity}, spectroscopy \cite{han2020time,spies2020terahertz,sterczewski2020terahertz} and other numerous applications. Because of the mentioned applications of THz waves, many THz components have been designed for manipulating THz waves, including absorbers \cite{huang2020active,xiong2020dual,huang2020broadband}, power splitters \cite{yang2017numerical,ung2012low,homes2007silicon,reichel2016broadband,hou2013terahertz}, modulators \cite{jakhar2020optically,chen2009metamaterial,jakhar2020integration}, and polarization power spitters \cite{zeng2019terahertz,lee2018broadband,berry2012broadband}. Among these THz components, THz power splitters (TPSs) are of critical importance to route THz waves. Homes et al. proposed a TPS based on silicon slab with several millimeters thickness for utilizing in spectroscopic measurements  in a broad THz range (0.2--10 THz) \cite{homes2007silicon}. In 2012, Ung et al. designed and fabricated a TPS by a thin silver sheet deposited on an ultra-thin low-density polyethylene plastic sheet. By changing the thickness of the silver sheet and utilizing its skin depth, they could achieve different splitting ratios. The splitting ratios achieved by transmission and reflection of the metalic-dielectric structure is in the frequency range of 0.5--1.5 THz \cite{ung2012low}. In 2013, Hou et al. proposed a TPS based on ferrite photonic crystals. They created line defects in photonic crystals to steer light and also embedded two silicon and ferrite rods at output branches to control various splitting ratios. By applying magnetic field to the ferrite rod, the refractive index of the ferrite rod changes and results in different splitting ratios \cite{hou2013terahertz}. In 2016, Reichel et al. designed and fabricated a TPS using T-junction metallic structures. They created metallic waveguides to route THz waves and a mechanical septum to achieve different splitting ratios over a frequency range of 0.15--0.3 THz \cite{reichel2016broadband}. In 2017, Yang et al. utilized temperature-dependence photonic crystals for realizing a 1 $\times$ 6 TPS at 1 THz. They used a multi-mode interferometer and four Y splitters that were created using line and point defects in photonic crystals \cite{yang2017numerical}. The mentioned TPSs suffer from need to external magnetic fields that enforce higher fabrication cost \cite{hou2013terahertz}, larger dimensions \cite{yang2017numerical,ung2012low,homes2007silicon,reichel2016broadband,hou2013terahertz}, inability for integration \cite{ung2012low,homes2007silicon,reichel2016broadband}, and working at a narrowband THz spectrum \cite{yang2017numerical}.

In this paper we present an all-dielectric compact TPS with ultra-small footprint and compatible with CMOS fabrication technology based on metamateials using inverse design. The inverse design uses particle swarm optimization (PSO) method to optimize the structure for realizing a 50:50 splitting ratio. In this device, we have used digitized structure of air holes in silicon slab deposited on a buried silicon dioxide substrate. The device contains one input and two output waveguides.

\begin{figure}[h]
\centering
\includegraphics[width=\linewidth]{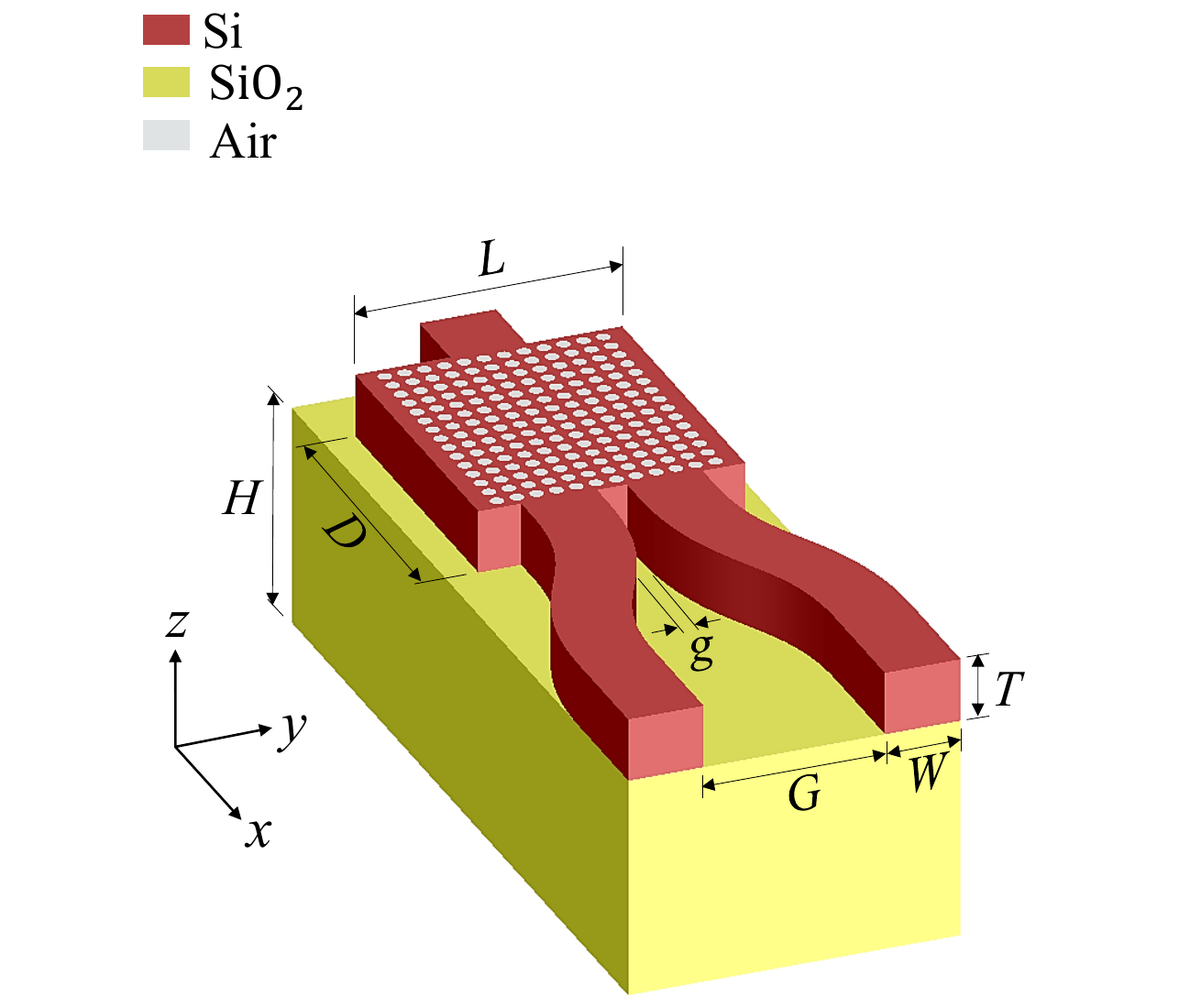}
\caption{Schematic of initial structure for the TPS.}
\label{fig:initial}
\end{figure}

For designing the device based on PSO method, an initial structure is needed. For this purpose, a digitized structure using etched air holes in Si slab on a SiO\textsubscript{2} substrate has been used as shown in Fig. \ref{fig:initial}. In this figure the radii of air holes and period of array are 9 and 26 \textmu m, respecticely. The parameters $L$, $D$, $H$, $G$, $g$, $W$, and $T$ are set to 350, 400, 1000, 237.6, 39.6, 100, and 100 \textmu m, respectively. The optimization region is a digitized structure with $L \times D \times T$ dimensions consisting of circular air holes in a silicon slab. In this structure, air holes are modeled as 1 and silicon is modeled as 0. IL is considered as figure of merit (FOM) for the proposed device that is minimized using PSO method.

For simulating and optimizing the structure, 3D finite-difference time-domain (FDTD) and PSO methods have been utilized. The generation size and maximum generation of the PSO method are 10 and 100, respectively. The mesh sizes for 3D FDTD are $dx= dy= dz=$ 3 \textmu m. Also, TE\textsubscript{00} polarization mode has been used to excite the waveguides. The electrical field component for TE\textsubscript{00} polarization mode is in $y$- direction. The working frequency range for the device is between 0.9--1.1 THz. In this range of frequency, the refractive index of Si and SiO\textsubscript{2} are 3.41 and 1.96, respectively. Schematic of initial structure for the 1 $\times$ 2 TPS is shown in Fig. \ref{fig:initial} for TE\textsubscript{00} polarization mode.

\begin{figure}[h]
\centering
\includegraphics[width=\linewidth]{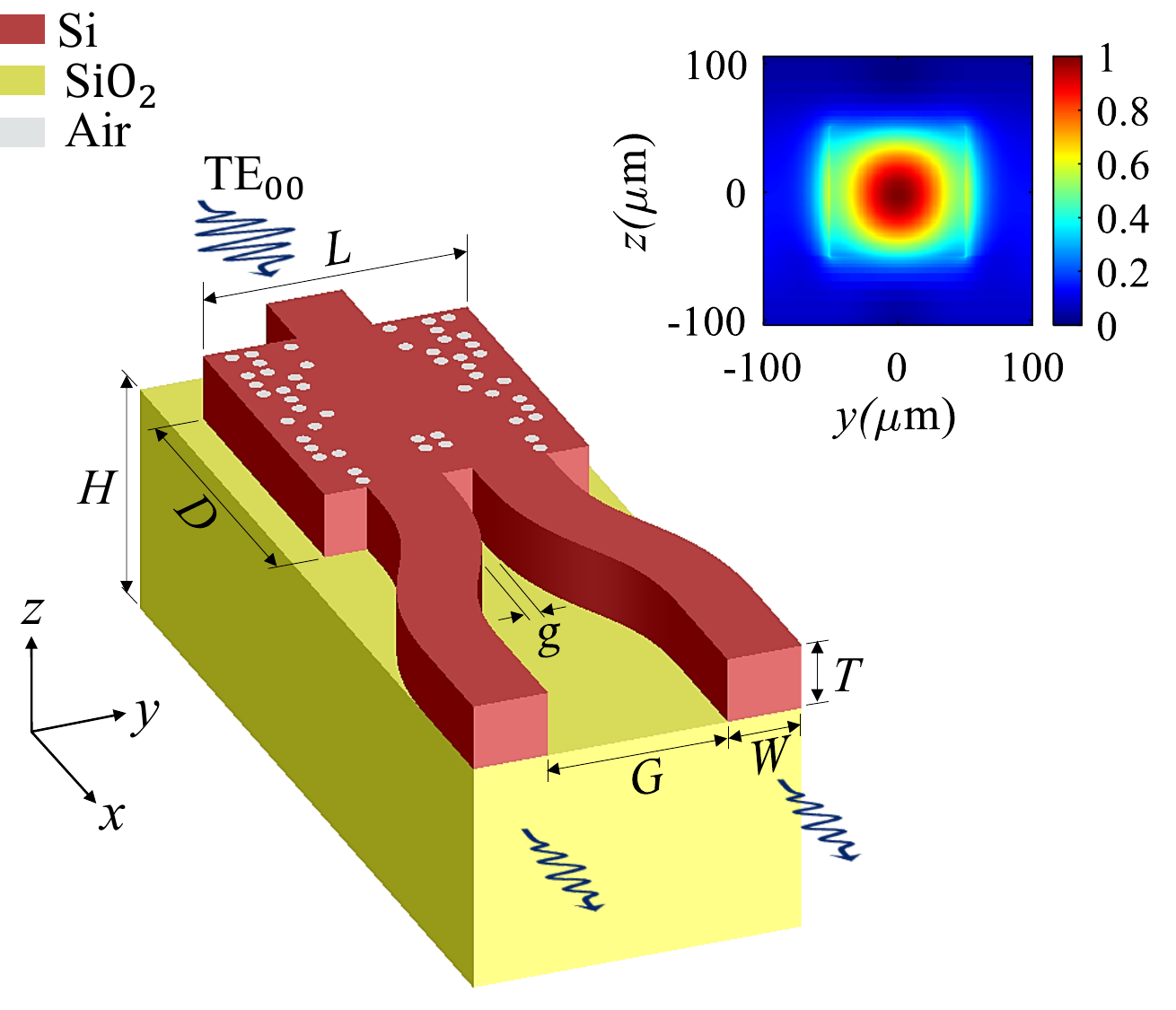}
\caption{Schematic of the optimized TPS structure.}
\label{fig:structure}
\end{figure}

Fig. \ref{fig:structure} represents the optimized structure by inverse design methos of PSO with a few circular air holes in silicon slab on a SiO\textsubscript{2} substrate. The electromagnetic power distribution of TE\textsubscript{00} incident polarization mode at cross section of the input waveguide has been depicted in the inset of Fig. \ref{fig:structure}.

\begin{figure}[h]
\centering
\includegraphics[width=\linewidth]{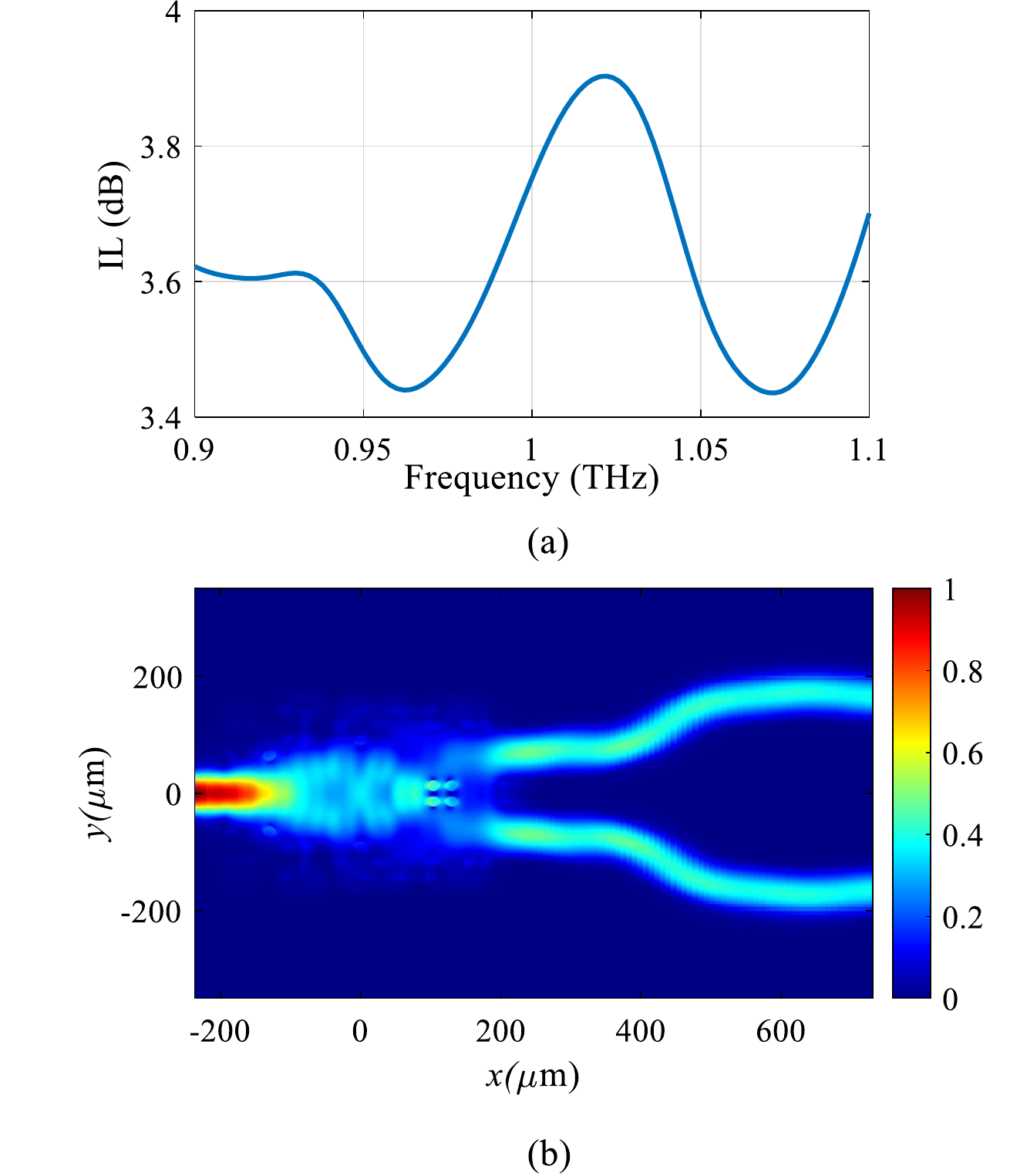}
\caption{(a) IL for incident wave with TE\textsubscript{00} mode, and (b) distribution of electromagnetic power over the structure at a wavelength of 1.55 \textmu m.}
\label{fig:fields}
\end{figure}

The air holes in the optimized structure act as scatterer points and based on interferences in the optimized region, constructive interferences take place and resulting in routing light to the output branches. Also, these holes make a photonic bandgap which prohibits THz waves to spread out in $y$- direction; so light can only be guided in the output branches. The lack of holes in the optimized region creates a guiding electromagnetic modes in the bandgap which results in steering the light to the outputs. In Fig. \ref{fig:fields}, the simulation results for electromagnetic power distribution and IL of the optimized 50:50 TPS structure have been shown. As demonstrated in Fig. \ref{fig:fields}(a), the IL is less than 3.9 dB that correspond to the efficiency roughly 81\% over the frequency range of 0.9--1.1 THz. Electromagnetic power distribution of the optimized TPS at the wavelength of 1.55 \textmu m is shown in Fig. \ref{fig:fields}(b) which represents the route of light over the device.

In this letter, we have proposed a 50:50 TE TPS based on digitized metamaterials. The device contains an input and two silicon waveguide outputs. It was optimized using PSO inverse design method. The optimized device represented IL of less than 3.9 dB over the frequency range of 0.9--1.1 THz. It contains a few air holes in silicon slab which are deposited on SiO\textsubscript{2} substrate. The advantages of the device contains compactness, efficient, small footprint, all-dielectric and compatible with CMOS fabrication technology. This device is a suitable candidate for high density THz circuits.

\begin{backmatter}
\bmsection{Acknowledgments} This work was supported by MSIT(Ministry of Science and ICT), Korea, under the ICT Creative Consilience program(IITP-2020-2011-1-00783) supervised by the IITP(Institute for Information \& communications Technology Planning \& Evaluation).
\end{backmatter}

\bibliography{References_OL}

\begin{thebibliography}{10}
\newcommand{\enquote}[1]{``#1''}

\bibitem{siegel2004terahertz}
P.~H. Siegel, {\protect\JournalTitle{IEEE transactions on microwave theory and
  techniques}} \textbf{52}, 2438 (2004).

\bibitem{brun2010terahertz}
M.-A. Brun, F.~Formanek, A.~Yasuda, M.~Sekine, N.~Ando, and Y.~Eishii,
  {\protect\JournalTitle{Physics in Medicine \& Biology}} \textbf{55}, 4615
  (2010).

\bibitem{smye2001interaction}
S.~Smye, J.~Chamberlain, A.~Fitzgerald, and E.~Berry,
  {\protect\JournalTitle{Physics in Medicine \& Biology}} \textbf{46}, R101
  (2001).

\bibitem{ergun2015terahertz}
S.~Ergun and S.~Sonmez, {\protect\JournalTitle{Journal of Management and
  Information Science}} \textbf{3}, 13 (2015).

\bibitem{cheng2020concealed}
Y.~Cheng, Y.~Wang, Y.~Niu, and Z.~Zhao, {\protect\JournalTitle{Optics express}}
  \textbf{28}, 6350 (2020).

\bibitem{harter2020generalized}
T.~Harter, C.~F{\"u}llner, J.~Kemal, S.~Ummethala, J.~Steinmann, M.~Brosi,
  J.~Hesler, E.~Br{\"u}ndermann, A.-S. M{\"u}ller, W.~Freude \emph{et~al.},
  {\protect\JournalTitle{Nature Photonics}} \textbf{14}, 601 (2020).

\bibitem{sarieddeen2020next}
H.~Sarieddeen, N.~Saeed, T.~Y. Al-Naffouri, and M.-S. Alouini,
  {\protect\JournalTitle{IEEE Communications Magazine}} \textbf{58}, 69 (2020).

\bibitem{petrov2020capacity}
V.~Petrov, D.~Moltchanov, Y.~Koucheryavy, and J.~M. Jornet,
  {\protect\JournalTitle{IEEE Transactions on Vehicular Technology}}
  \textbf{69}, 6822 (2020).

\bibitem{han2020time}
P.~Han, X.~Wang, and Y.~Zhang, {\protect\JournalTitle{Advanced Optical
  Materials}} \textbf{8}, 1900533 (2020).

\bibitem{spies2020terahertz}
J.~A. Spies, J.~Neu, U.~T. Tayvah, M.~D. Capobianco, B.~Pattengale, S.~Ostresh,
  and C.~A. Schmuttenmaer, {\protect\JournalTitle{The Journal of Physical
  Chemistry C}} \textbf{124}, 22335 (2020).

\bibitem{sterczewski2020terahertz}
L.~A. Sterczewski, J.~Westberg, Y.~Yang, D.~Burghoff, J.~Reno, Q.~Hu, and
  G.~Wysocki, {\protect\JournalTitle{ACS Photonics}} \textbf{7}, 1082 (2020).

\bibitem{huang2020active}
J.~Huang, J.~Li, Y.~Yang, J.~Li, Y.~Zhang, J.~Yao \emph{et~al.},
  {\protect\JournalTitle{Optics express}} \textbf{28}, 7018 (2020).

\bibitem{xiong2020dual}
H.~Xiong, Q.~Ji, T.~Bashir, and F.~Yang, {\protect\JournalTitle{Optics
  express}} \textbf{28}, 13884 (2020).

\bibitem{huang2020broadband}
J.~Huang, J.~Li, Y.~Yang, J.~Li, J.~Li, Y.~Zhang, and J.~Yao,
  {\protect\JournalTitle{Optics Express}} \textbf{28}, 17832 (2020).

\bibitem{yang2017numerical}
L.~Yang and L.~Jiu-Sheng, {\protect\JournalTitle{Journal of the European
  Optical Society-Rapid Publications}} \textbf{13}, 1 (2017).

\bibitem{ung2012low}
B.~S.-Y. Ung, C.~Fumeaux, H.~Lin, B.~M. Fischer, B.~W.-H. Ng, and D.~Abbott,
  {\protect\JournalTitle{Optics express}} \textbf{20}, 4968 (2012).

\bibitem{homes2007silicon}
C.~C. Homes, G.~L. Carr, R.~P. Lobo, J.~D. LaVeigne, and D.~B. Tanner,
  {\protect\JournalTitle{Applied optics}} \textbf{46}, 7884 (2007).

\bibitem{reichel2016broadband}
K.~S. Reichel, R.~Mendis, and D.~M. Mittleman,
  {\protect\JournalTitle{Scientific reports}} \textbf{6}, 28925 (2016).

\bibitem{hou2013terahertz}
Y.~Hou, F.~Fan, X.-H. Wang, and S.-J. Chang, {\protect\JournalTitle{Optik}}
  \textbf{124}, 5285 (2013).

\bibitem{jakhar2020optically}
A.~Jakhar, P.~Kumar, A.~Moudgil, V.~Dhyani, and S.~Das,
  {\protect\JournalTitle{Advanced Optical Materials}} \textbf{8}, 1901714
  (2020).

\bibitem{chen2009metamaterial}
H.-T. Chen, W.~J. Padilla, M.~J. Cich, A.~K. Azad, R.~D. Averitt, and A.~J.
  Taylor, {\protect\JournalTitle{Nature photonics}} \textbf{3}, 148 (2009).

\bibitem{jakhar2020integration}
A.~Jakhar, P.~Kumar, S.~Husain, V.~Dhyani, and S.~Das,
  {\protect\JournalTitle{ACS Applied Nano Materials}} \textbf{3}, 10767 (2020).

\bibitem{zeng2019terahertz}
H.~Zeng, Y.~Zhang, F.~Lan, S.~Liang, L.~Wang, T.~Song, T.~Zhang, Z.~Shi,
  Z.~Yang, X.~Kang \emph{et~al.}, {\protect\JournalTitle{IEEE Transactions on
  Terahertz Science and Technology}} \textbf{9}, 491 (2019).

\bibitem{lee2018broadband}
W.~S. Lee, S.~Nirantar, D.~Headland, M.~Bhaskaran, S.~Sriram, C.~Fumeaux, and
  W.~Withayachumnankul, {\protect\JournalTitle{Advanced Optical Materials}}
  \textbf{6}, 1700852 (2018).

\bibitem{berry2012broadband}
C.~W. Berry and M.~Jarrahi, {\protect\JournalTitle{Journal of Infrared,
  Millimeter, and Terahertz Waves}} \textbf{33}, 127 (2012).

\end{thebibliography}

\end{document}